\documentclass[superscriptaddress, twocolumn, prb, bibnotes, aps, longbibliography, floatfix]{revtex4-2}
\usepackage{comment, siunitx}
\usepackage[utf8]{inputenc}
\usepackage{graphicx}
\usepackage{afterpage}
\usepackage{amsmath, amssymb, bbm, bm}
\usepackage{xcolor}
\usepackage{float,placeins}

\begin{document}

\title{Single-shot measurement of photonic topological invariant}

\author{Nathan Roberts}
\affiliation{Department of Physics, University of Bath, Claverton Down, Bath BA2 7AY, UK}
\affiliation{Centre for Photonics and Photonic Materials, University of Bath, Bath BA2 7AY, UK}

\author{Guido Baardink}

\affiliation{Department of Physics, University of Bath, Claverton Down, Bath BA2 7AY, UK}

\author{Anton Souslov}
\email{as3546@cam.ac.uk}
\affiliation{Department of Physics, University of Bath, Claverton Down, Bath BA2 7AY, UK}
\affiliation{TCM Group, Cavendish Laboratory, JJ Thomson Avenue,
Cambridge, CB3 0HE, UK}

\author{Peter J.~Mosley}
\email{p.mosley@bath.ac.uk}
\affiliation{Department of Physics, University of Bath, Claverton Down, Bath BA2 7AY, UK}
\affiliation{Centre for Photonics and Photonic Materials, University of Bath, Bath BA2 7AY, UK}

\begin{abstract}
Topological design enables robustness to be engineered into a system. 
However, a general challenge remains to experimentally characterize topological properties. In this work, we demonstrate a technique for directly observing a winding-number invariant using a single measurement. By propagating light with a sufficiently broad spectrum along a topological photonic crystal fiber, we calculate the winding number invariant from the output intensity pattern. We quantify the capabilities of this single-shot method, which works even for surprisingly narrow and asymmetric spectral distributions. 
We demonstrate our approach using topological fiber, but our method is generalizable to other platforms.  Our method is experimentally straightforward: we use only a broadband input excitation and a single output to measure the topological invariant. 
\end{abstract}
\maketitle
\section{Introduction}
Topological band theory connects a system's physical properties, such as wave guidance of light and sound, to robust integer invariants~\cite{Hasan2010, Ozawa2019, Xue2022}. 
Topological waveguides use this connection to tie experimental observables to an environmentally independent topological band gap. 
The topology then endows the observable properties with enhanced robustness to changes in geometry, such as imperfections introduced during fabrication~\cite{Wang2009}. These advances promise to protect photonic~\cite{Khanikaev2013, Shalaev2019,Hafezi2013,Rechtsman2013}, acoustic~\cite{Yang2015,Khanikaev2015, Souslov2017}, and delicate quantum states~\cite{blanco-redondo2018, Wang2019, Ren2021,Duenas2021,Yan2021,Blanco-Redondo2020} from physical disorder. The resulting designs and realizations of resilient topological systems~\cite{Yang2015, Souslov2017, Khanikaev2013, Shalaev2019,Ren2021, Wang2008, Wang2009, blanco-redondo2018, Khanikaev2013, Raghu2008, Hafezi2013,Rechtsman2013,Wang2019,Duenas2021,Yan2021,Abbaszadeh2020,Khanikaev2015,Bertoldi2017,Blanco-Redondo2020} are based on a variety of materials, geometries, and topological invariants. However, an outstanding challenge across these implementations is that they are difficult to characterize experimentally. 

Finite systems characterized by nontrivial topological invariants can feature edge-localized states that are predicted by bulk-boundary correspondence~\cite{Halperin1982, Hatsugai1993, Qi2006}. Whenever an invariant changes, a topological band gap locally closes and reopens. This closing of a band gap gives rise to a localized boundary mode that lives on the system's edge, regardless of shape or deformation.
Although observing bulk-boundary modes suggests the presence of topology, this observation is insufficient for topological classification. For example, different topological invariants induce edge modes robust to different types of disorder, and correctly identifying the invariant helps to classify the degree of robustness~\cite{Qi2011}. In addition, edge modes are not exclusive to topological systems~\cite{Mondragon-Shem2014, Scollon2020, Ardakani1977}. Imperfection-localized modes can occur at defects near the boundary of non-topological systems. As a final example, topological edge modes always have an associated width, called the penetration depth, that describes how far the mode reaches into the bulk~\cite{Kane2014}. In the case of a large penetration depth and small system size, the distinction of topological edge modes from bulk modes becomes blurred. These examples demonstrate the need for an accessible way to characterize topology by directly measuring the topological invariant.

Classifying topology within 2+1D waveguide systems is especially difficult because one cannot straight-forwardly observe topological bands of supported eigenmodes. In condensed matter systems or in 2D photonic crystals, topological band gaps are characterized using frequency-resolved measurements.
However, in the case of an optical waveguide, the topology is connected to modal propagation constants which cannot be readily observed. 
Although we cannot measure the full band structure, we show how to distinguish topological states through differences in light propagation in the bulk of topological materials.

In the case of topological winding numbers, one can turn to direct observations of the invariant~\cite{Rudner2009,Atala2013,Longhi2013,Wang2019,Zeuner2015,Cardano2017,Maffei2018}. Reference~\cite{Meier2016} shows that for atomic wires with quenched dynamics, the integrated absorption images can distinguish between topological and trivial systems in the bulk. Analogously, Refs.~\cite{Wang2019, Zeuner2015, Cardano2017,Maffei2018} demonstrate direct measurements of photonic winding numbers where intensity is measured as a function of propagation length. In Ref.~\cite{Wang2019}, the experimental difficulty of this method is highlighted, requiring the fabrication of many waveguide arrays of various lengths. In contrast, we develop, perform, and evaluate an experimental method to measure the topological winding number from a single sample using a single-shot measurement. 
The speed and simplicity of our method enable real-time measurement of topology, which could be crucial for understanding how a system's topology changes with rapid deformation or distortion.  
\begin{figure}[tbp]
\centering
\includegraphics[width=\linewidth]{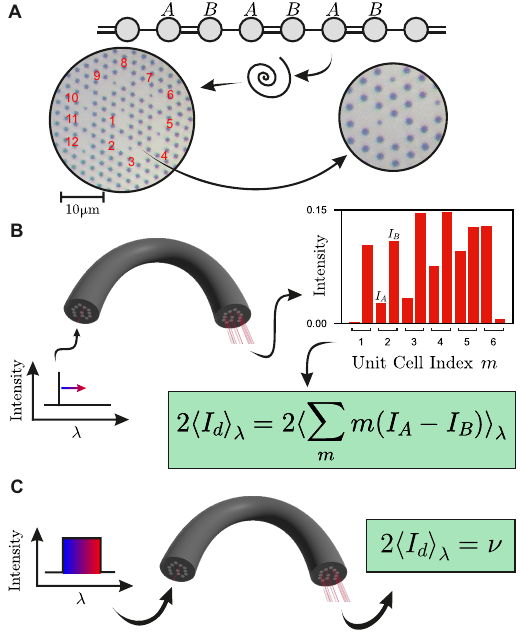}
\caption{Schematic of the single-shot measurement. \textbf{A}, Cross-section of the 12-core topological photonic crystal fiber based on the SSH chain (left). The  different air hole sizes 
 (see right zoom) give rise to alternating weak--strong inter-core couplings. \textbf{B}, A  narrow spectrum of light is coupled into a bulk core (left). The resultant intensity profile (right) is used to calculate a weighted intensity difference $I_d$, which is then wavelength-averaged to measure the system's topological invariant. 
\textbf{C}, Our method uses a 
broadband spectrum, allowing the invariant to be calculated in a single shot.}
\label{fig:1}
\end{figure}

We demonstrate our method using a topological photonic crystal fiber containing 12 coupled light-guiding cores, shown in Fig.~\ref{fig:1}A~\cite{Roberts2022}. The schematic above the microscope image shows gray circles as fiber cores and black lines as alternating coupling strengths, corresponding to the topological model  of a Su-Schrieffer-Heeger (SSH) chain~\cite{Su1979}. 
To experimentally measure the invariant in this fiber, we excite a single bulk core using a broad spectrum of light.
At the output of the fiber, each wavelength gives a weighted intensity difference between the two SSH chain sublattices (see Fig.~\ref{fig:1}B). When averaging over the broad spectrum, the weighted intensity difference can be shown to give a winding number measurement. Therefore, this single-shot experiment characterizes the topology by having the fiber perform the averaging for us, as shown in Figure~\ref{fig:1}C. 

Given the generality of our method, this technique can be translated to any platform exhibiting a photonic SSH invariant, from on-chip laser-written waveguides to bulk optics. 
Additionally, it can be broadened to measure winding numbers in 2D SSH chains~\cite{Lu:20}, or $\textrm{SSH}_{4}$ chains~\cite{Maffei2018}. Conceptually, our method can be applied to any wave-guiding medium, enabling direct, single-shot observation of topological invariants in acoustic, mechanical, or condensed matter implementations.
This offers an alternative to measuring topology as a function of discrete momenta, which has seen success in etched semiconductor structures~\cite{StJean2021}.
With all of these domain specific applications in mind, we envision a future toolkit comprised of advanced characterization techniques such as ours, which will be used in the evaluation of topologically enhanced devices.

\section{Theory}

Light propagation in fiber is governed by the scalar wave equation: 
\begin{equation}\label{eq:scalarwaveeq}
    \nabla_{\perp}^{2}\psi(x,y) + 
    k_0^2 n^2 \psi(x,y) = \beta^2 \psi(x,y),
\end{equation}
where $\nabla_{\perp}^{2} \equiv \partial_x^2 + \partial_y^2$ is the transverse Laplacian and $k_0 = 2 \pi/\lambda$ is the free-space wavenumber, $n$ is the refractive index of the glass, $\psi$ is the electric field $E_{x}$ or $E_{y}$, and $\beta$ is the propagation constant (the $z$ component of the wave vector). To find the supermodes of a weakly coupled multicore fiber, we use the eigenvalue equation $\rm{C}\textbf{u}=$$\Delta \beta\textbf{u}$, where $\Delta \beta$ is the change in propagation constant for a supermode relative to the propagation constant of a single core (where entries in matrix $\rm{C}$ depend on $A-B$ couplings $C_1$ within a unit cell and couplings $C_2$ between unit cells). By inducing alternating coupling strengths within and between unit cells, a topological Su-Schrieffer-Heeger (SSH) chain is created~\cite{Roberts2022}. This chain supports edge modes that are robust to symmetry-preserving disorder and is characterized by a topological winding number. In this system, the winding number describes how the phase difference within a unit cell varies as a function of the phase difference between unit cells. If the path taken by the phase variation encloses the origin, the winding number is one and the system will feature topological edge states. Otherwise the fiber is considered topologically trivial.  
\begin{figure*}[ht]
\centering
\includegraphics[width=\linewidth]{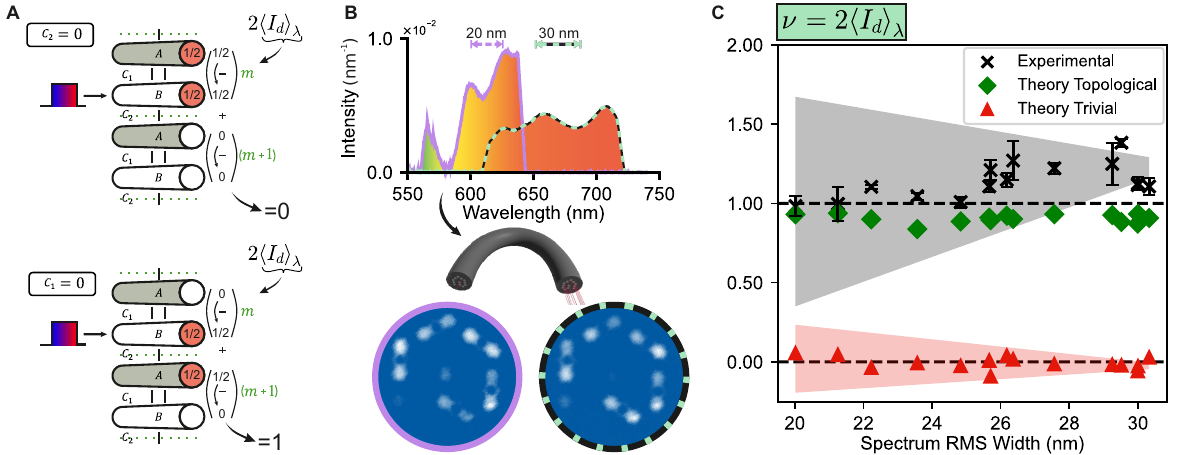}
\caption{\textbf{A} Heuristic explanation of the connection between the output intensity profile and the topological invariant. A broad spectrum excites a single core and the output intensity profiles are considered for two extreme cases, $C_1 = 0$ and $C_2 = 0$. For the topologically trivial case $C_2 = 0$, light stays within a single unit cell and the average intensity difference is zero. 
For the non-trivial case $C_1 = 0$, light cannot couple to the other core within the same unit cell. On average, half the light intensity ends up in the neighboring unit cell, making the weighted intensity difference $2\langle I_d \rangle_\lambda = 1$.
\textbf{B} Schematic explanation of our experiment. The intensity distributions per unit wavelength for both the narrowest spectrum (purple) and the widest spectrum (dashed teal) are shown in the plot. The intensity distributions are used to excite core six of the topological fiber before the output is imaged onto a camera. The two intensity plots shown correspond to the two spectra shown in the plot. \textbf{C} Experimental data showing the effects of changing the spectral width on the winding-number measurement ($\nu$). The black crosses are experimental averages of three measurements, with the error bars being their standard deviation. The observed winding numbers stay around the expected value of one, but the uncertainty associated with the measurements (gray shaded region) grows as the root mean square (RMS) spectral width decreases. The green diamonds and red triangles show the theoretical predictions of our measurement when the experimental spectra are propagated. The diamonds correspond to the system's topological state with the same couplings as in our experiment, while for the triangles, the $C_1$ and $C_2$ couplings are flipped, leaving the system in a topologically trivial phase.}
\label{fig:2}
\end{figure*}

Reference~\cite{Roberts2022} introduced a topological fiber which we now use to demonstrate our single-shot measurement technique. 
To measure the winding number, $\nu$, we inject a broadband spectrum into a single core of our topological multicore fiber and observe the output profile across all cores. To find $\nu$, we first define the weighted intensity difference $I_{d} = \sum_m m(I_{Am} - I_{Bm})$, also known as the mean chiral displacement~\cite{Maffei2018}, or photon population difference center~\cite{Wang2019}, to be the output intensity difference between the two cores ($I_{Am}$ and $I_{Bm}$) within one unit cell, weighted by the unit cell labeled $m$ and summed over all unit cells to give a single value at a given wavelength $\lambda$.
As shown in Figure~\ref{fig:1}C, Eq.~\ref{eq:id_lambda}, and detailed in the supplemental material (SM), the measurement of $I_{d}$ across a broad range of wavelengths $\lambda$ gives us a direct measurement of $\nu$: $\nu = 2\langle I_{d} (\lambda) \rangle_{\lambda}$,

\begin{align}\label{eq:id_lambda}
 \int_{\lambda_a}^{\lambda_b} &I_\textrm{d}(z) g(\lambda)\, d\lambda =
 \frac{\nu}{2} + \\
 &\frac{1}{4\pi}\int_0^{2\pi} \int_{\lambda_a}^{\lambda_b} g(\lambda) \cos(2|\zeta|z)\, d\lambda \frac{d \phi}{d \theta} d\theta \nonumber,
\end{align}
where $\zeta = C_{1} + C_{2}e^{i\theta}$, $\phi = \arg \zeta(\theta)$, $g(\lambda)$ is the wavelength distribution in the illuminating spectrum.

In Eq.~(\ref{eq:id_lambda}), we show that averaging the weighted intensity difference over a distribution specified by $g(\lambda)$, returns the winding number if the oscillatory term is zero. The coupling strengths $C_{1,2}$ are well approximated as linear over our regime of interest, so we can express $|\zeta(\lambda)| = f(\theta) + h(\theta)\lambda$ and write the whole oscillatory term as a product of nondimensionalized complex exponentials:

\begin{align}
\frac{1}{8\pi}\int_0^{2\pi}\tau(\theta) e^{i2f(\theta)z} \int_{P_a}^{P_b} G\left(\frac{P}{\sigma'}\right)e^{iP} dP\frac{d \phi}{d \theta} d\theta 
    + \mathrm{c.c.}
\end{align}
To nondimensionalize, we make the substitutions $P = 2h(\theta)\lambda z$, $\tau(\theta) = 1/(2h(\theta)z)$, and $d\lambda = \tau(\theta) dP$. We also scale the normalised weighting function $g(\lambda)$ by a characteristic width $\sigma$, $ G(\lambda/\sigma) = G(P/\sigma')$, defining a new function $G(P/\sigma')$, where  $\sigma' = 2\sigma h(\theta) z$. 
If the oscillations in $e^{iP}$ occur over a shorter period than the changes in the spectral distribution $G(P/\sigma')$, the stationary phase approximation applies and $\int_{P_a}^{P_b} G(\frac{P}{\sigma'})e^{iP} dP \rightarrow 0$. As the integral approaches zero, the oscillatory term becomes small compared to the integrated weighted intensity difference [left-hand side Eq.~(\ref{eq:id_lambda})], causing the last term in Eq.~(\ref{eq:id_lambda}) to vanish. 
The only remaining contribution to the winding number in Eq.~(\ref{eq:id_lambda}) comes from the integrated weighted intensity difference $\nu = 2\langle I_d(\lambda)\rangle_\lambda$. 

Heuristically, this correspondence between light intensity and topology is shown in Fig.~\ref{fig:2}A and can be understood by considering two topologically obstructed cases: \emph{(i)} $C_1 = 0$ and \emph{(ii)} $C_2 = 0$. \emph{(i)} When $C_1 = 0$, half of the light couples into the unit cell neighbouring the injection site, leading to $I_{A1} - I_{B1}  = 1/2$ (on average) and $\nu = 1$. \emph{(ii)} By contrast, for $C_2 = 0$, all of the light is split evenly within the (same) unit cell of the injection site, leading to $I_{Am} - I_{Bm}  = 0$ (on average, for all $m$) and $\nu = 0$.
\begin{figure*}[t]
\centering
\includegraphics[width=\linewidth]{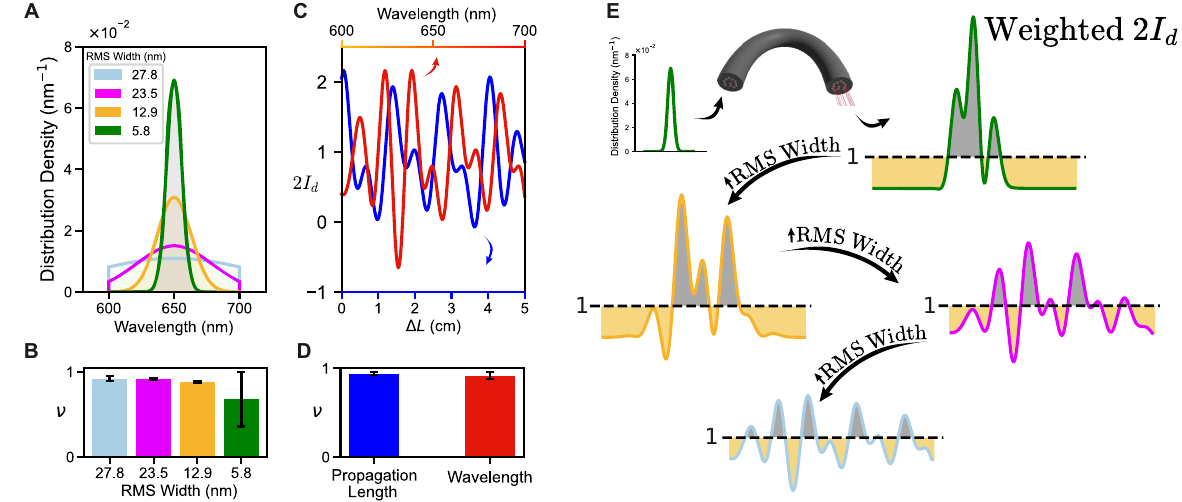}
\caption{\textbf{A}, Four example distributions of the input spectrum. We vary the root mean square (RMS) width of the input spectrum from \SI{27.8}{nm} (Turquoise) to \SI{5.8}{nm} (dark green) by reducing the standard deviation of the distribution.  \textbf{B}, Shows the calculated winding number ($\nu$) for each of these input spectra as a function of the RMS width of the distribution. \textbf{C}, Response of the weighted intensity difference to changing wavelength (red) and changing distance (blue) in the topologically non-trivial case. Both plots show twice the weighted intensity difference oscillating around one, the expected value of the winding number that characterizes the system. \textbf{D}, Winding numbers calculated by averaging the wavelength and distance curves plotted in \textbf{C}. \textbf{E} Product of the distribution density and $2I_d$ (which approaches the winding number as shown in Eq.~\ref{eq:id_lambda}), for the distributions shown in \textbf{A}. We show graphically that the mean of this function becomes closer to the winding number, $\nu =1$ as the RMS width of the exciting spectrum increases.}
\label{fig:3}
\end{figure*} 
\section{Results and Discussion}

Experimentally, we demonstrate our method by illuminating core 6 (see Fig.~1) with a variety of broadband spectra. We use a supercontinuum generating fiber and a wavelength filter built into a folded 4-f line, to control the bandwidth of the light source. Following the 4-f line, we couple light from the broadband source into an endlessly single mode photonic crystal fibre which is butt-coupled to our topological photonic crystal fiber. The output face is imaged onto a camera as shown in Fig.~\ref{fig:2}B. In Fig.~\ref{fig:2}C, we plot the average observed winding number as a function of the root mean square (RMS) width of the illuminating spectra (defined as $\sqrt{\sum_i (\lambda_{i}-\Tilde{\lambda})^2 P(\lambda_i)}$, where $P(\lambda_i)$ is the normalised intensity at a given wavelength $\lambda_i$ and $\Tilde{\lambda}$ is the distribution's mean wavelength). Two extreme examples of input spectra are shown in Fig.~\ref{fig:2}B, the purple line corresponds to the narrowest and least uniform spectrum, whereas the dotted teal and black line shows the broadest spectrum. We show that for these spectra, the calculated winding number remains close to the expected value of $ 
\nu = 1$. The speed of our measurement makes repeated observations straightforward, enabling experimental uncertainty to be quantified. 
We plot the average winding number and the standard deviation in Fig.~\ref{fig:2}C. We find good agreement between the observed winding number and our system's topological classification. However, this agreement obscures a crucial source of  uncertainty, introduced via the unweighted intensity difference and shown as the shaded gray and red regions in Fig.~\ref{fig:2}C. 

To understand the origin of this uncertainty, we must consider the labelling of unit cells in the weighted intensity difference calculation. As shown in SI, the system's winding number is equal to twice the observed weighted intensity difference when the injection unit cell, $m^{\star}$, is labelled by $m^{\star}=0$. This is a direct consequence of the more general relation: $\nu = 2 I_{d}- 2 m^{\star}\sum_{m} (I_{Am} - I_{Bm})$. This relation states that the measured value of the weighted intensity difference ($I_d$) is shifted by $2m^{\star}$ times the unweighted intensity difference, which is defined as the difference in intensity between A and B cores, summed over all unit cells. In the case of a perfect fiber and flat spectrum, the unweighted intensity difference should be zero, as there would be no preference for light to be in either the A or B sites. However, in the experiment, the unweighted intensity difference can be non-zero due to imperfections in the fiber structure.

While the choice $m^{\star} = 0$ is sufficient for the measurement to result in $\nu = 1$ across all our input spectra, we want to observe the same winding number regardless of unit cell label. We consider labelling the unit cell with an alternative established convention: labelling the unit cell at the inner edge to be $m = 1$ and counting outwards so that $m^{\star} = 3$~\cite{Roberts2022,Wang2019}.
We use the difference between the two labelling conventions $m^{\star} = 3$ and $m^{\star} = 0$ to highlight the labelling related uncertainty. 
We linearly fit the range of uncertainties to estimate the overall uncertainty as the gray shaded region in Fig.~\ref{fig:2}C. 
While our measurement returns a non-zero winding number for all spectral widths, the uncertainty prohibits the value from being well defined for spectral widths below \SI{24}{nm}.
 To understand why this method, which is initially derived considering a perfectly flat spectrum, can still measure the invariant in the presence of imperfect and narrow spectra, we turn to the analytic theory underpinning this technique.

In Fig.~\ref{fig:3}, we use analytic theory and numerical simulations to understand the limits of our single-shot measurement. Figure~\ref{fig:3}A shows four spectral distributions that are used as the input for our analytic model. We use a distribution which is zero at wavelengths outside the region of \SI{600}{nm} to \SI{700}{nm} and which is otherwise a Gaussian  with standard deviations $\sigma$ of: \SI{64.6}{nm}, \SI{28.9}{nm}, \SI{12.9}{nm}, and \SI{5.8}{nm} as inputs to our model fiber with a \SI{23.7}{cm} fiber length to mirror our experiment. Figure~\ref{fig:3}B shows the measured winding number for each input distribution. As expected from the experiment, the observed winding number is calculated to be $\nu \approx 1$ for the three widest distributions, before falling off as the spectrum narrows. This drop-off begins around $\sigma = \SI{12.9}{nm}$ and as the RMS spectral width reaches $\sigma = \SI{5.8}{nm}$, the measurement no longer returns a value close to one, and our measurement technique breaks down. To understand this breakdown, we plot twice the wavelength-dependent weighted intensity difference [$2I_d(\lambda)$, which should average to $\nu$] in Fig.~\ref{fig:3}C. 
The red curve shows how this weighted intensity difference oscillates around the winding number, with an average peak-to-peak distance of $p = \SI{16}{nm}$. 
If the input spectrum becomes narrower than this width, that is $\sigma < p$, the wavelength average will be unevenly weighting parts of $2I_d(\lambda)$. Therefore, for narrow spectra, the single-shot technique will not measure the winding number.
Thus, our method requires a broadband spectrum, with  $\sigma > p$, where $p$ can be computed numerically \emph{a priori} for a given experimental platform. In particular, $p$ depends on the wavelength-dependent coupling strength between the neighboring cores; hence the parameters of the single-shot measurement can be optimized for a given system. 

The requirement of a sufficiently broad spectrum makes sense both within the context of our technique, and when considering previous work observing topological invariants. In the context of quenched atomic dynamics, the topological invariant arises out of the long-time average of the measured absorption~\cite{Meier2016,StJean2021}.
The long-time average in atomic dynamics translates to a long-distance average for measurements of the topological invariant in waveguides~\cite{Wang2019,Maffei2018,Cardano2017}.
For our system, light couples between cores as a function of distance, changing the output intensity profile as it propagates. Equivalently, we can change the final intensity profile for a given length by changing the coupling strengths. This is experimentally straightforward because the coupling strength is linearly dependent on wavelength (over our range of interest), so we can observe changes equivalent to propagating different distances by changing the wavelength of the light.
While we demonstrate our method using centimeters of fiber, by scaling the coupling strengths appropriately, one could make a direct observation of a topological invariant over an arbitrary length.

Using numerics, we compare the single-shot measurement to other characterization techniques. In Fig.~\ref{fig:3}C, the blue curve shows the variation $2I_d(\Delta L)$ of (twice) the weighted intensity difference as a function of fiber length. This curve oscillates around its winding-number average for one given input wavelength, $\lambda = \SI{650}{nm}$. To experimentally observe the winding number using this previous method, 
it would be necessary to compare propagation in the same fiber, but cut to different lengths~\cite{Wang2019}, or instead to image the scattering of light perpendicular to the direction of propagation over the entire length of the device~\cite{Xu2022}. In contrast with Ref.~\cite{Xu2022}, our single-shot method only requires access to the input and output of the waveguide in order to measure the topological invariant. In Fig.~\ref{fig:3}C and D, we confirm numerically that the length and wavelength-based measurements both return $2I_d$ curves that oscillate around the topological invariant $\nu$, successfully benchmarking our single-shot measurement against previous approaches.

In conclusion, we have demonstrated a single-shot measurement technique for calculating the winding number of a topological system. Despite an initial assumption of a flat, broadband excitation spectrum, we find that our method distinguishes topological states using initial excitation spectra that are  narrow and asymmetric. Using numerical simulations and analytic theory, we find the spectral robustness of our method is well explained using the stationary phase approximation. To characterize the topology of a given system, we can use our results to define an optimal combination of measurement parameters, such as spectral width, waveguide length, and coupling strength. We envision our method to be highly generalizable to topological systems beyond the fiber-optic platform in this work.

\subsection*{Acknowledgements}
This work is supported by the Air Force Office of Scientific Research under award number FA865522-1-7028. A.S.~acknowledges the support of the Engineering and Physical Sciences Research Council (EPSRC) through New Investigator Award No.~EP/T000961/1 and of the Royal Society under grant No.~RGS/R2/202135.

 \subsection*{Data Availability}
Raw experimental data are available on Zenodo at DOI:10.5281/zenodo.10534852 under an MIT license (Ref.~\cite{roberts_2024_10534852}).
\subsection*{Competing Interests}
 The authors declare that they have no competing interests.
\bibliographystyle{apsrev4-1}

\end{document}